# A theory of consciousness: computation, algorithm, and neurobiological realization

J. H. van Hateren

Johann Bernoulli Institute for Mathematics and Computer Science,
University of Groningen, The Netherlands; j.h.van.hateren@rug.nl

**Abstract** The most enigmatic aspect of consciousness is the fact that it is felt, as a subjective sensation. The theory proposed here aims to explain this particular aspect. The theory encompasses both the computation that is presumably involved and the way in which that computation may be realized in the brain's neurobiology. It is assumed that the brain makes an internal estimate of an individual's own evolutionary fitness, which can be shown to produce a special, distinct form of causation. Communicating components of the fitness estimate (either for external or internal use) requires inverting them. Such inversion can be performed by the thalamocortical feedback loop in the mammalian brain, if that loop is operating in a switched, dual-stage mode. A first (nonconscious) stage produces forward estimates, whereas the second (conscious) stage inverts those estimates. It is argued that inversion produces another special, distinct form of causation, which is spatially localized and is plausibly sensed as the feeling of consciousness.

## 1 Introduction

The terms 'consciousness' and 'conscious' have various meanings. They may refer to the state of being awake (as in 'regaining consciousness'), the process of gaining access to certain facts as they affect the senses or are retrieved from memory (as in 'becoming conscious of something'), and the subjective sensation associated with experiencing (e.g., when feeling pain or joy, and when undergoing a visual experience). The primary topic of this article is the latter meaning, sometimes referred to as phenomenal consciousness (Block 1995). The main purpose here is to explain why consciousness is *felt*. Nevertheless, the explanation given below has implications for the first two meanings as well.

The science of consciousness is making considerable progress by studying the neural correlates of consciousness (Dehaene 2014; Koch et al. 2016). However, these studies primarily aim to identify which particular neural circuits are involved in consciousness, but not how and why exactly such neural mechanisms would produce subjective experience. Theories that explicitly address the latter typically focus on a specific neural, cognitive, or informational process, which is then hypothesized to be accompanied by consciousness. There is no shortage of such proposals, more than could be mentioned here. Some representative examples are: a narrative that the brain compiles from competing micro-narratives (Dennett 1991); a regular, but unspecified physiological process (Searle 2013); broadcasting messages to a widely accessible global workspace within the brain (Baars 1988); neuronal broadcasts to a global neuronal workspace (Dehaene et al. 2003); having representations in the form of trajectories in activity space (Fekete and Edelman 2011); the self perceiving its own emotional state (Damasio 1999); having representations about representations (Lau and Rosenthal 2011); attending to representations (Prinz 2012); perceiving socially observed attention (Graziano and Kastner 2011); recurrent neuronal processing (Lamme and Roelfsema 2000); having a dynamic core of functional neural clusters (Edelman and Tononi 2000); having the capacity to integrate information (Oizumi et al. 2014); and having unified internal sensory maps (Feinberg and Mallatt 2016).

Only some of these theories are closely associated with neurobiological measurements. In particular, the global neuronal workspace theory (Dehaene et al. 2003) assumes that, when perceiving, the brain first engages in a nonconscious processing stage, which can (but need not) lead to a global second stage (a brain-wide 'ignition'). There is empirical support for consciousness arising in the second stage, which is taken to provide a globally accessible workspace for the results of the first stage, as well as for subsequent processing. Another well-known neural theory (Lamme and Roelfsema 2000) also assumes two subsequent stages. When a visual stimulus is presented, there is first a fast forward sweep of processing, proceeding through the cortex. This first stage is nonconscious. Only a second stage of recurrent processing, when earlier parts of the visual cortex are activated once more by feedback from later parts, is taken to be conscious (with, again, empirical support). Finally, Edelman and Tononi (2000) propose that a loop connecting thalamus and cortex forms a dynamic core of functional neural clusters, varying over time. This core is assumed to integrate and differentiate information in such a way that consciousness results (for an elaborate theory along these lines see Oizumi et al. 2014).

The theory of consciousness that is proposed in the present study takes a somewhat unusual approach, as it first constructs a stochastic causal mechanism that plausibly produces something distinct that may be experienced. Only then does it conjecture which neural circuits in the brain are good candidates for the mechanism's implementation, and how that could be tested. The approach bears resemblance to the research strategy that was proposed by Marr (1982) for the visual system. It assumes that it is useful to distinguish three levels of explanation when analysing a complex system such as the brain. At the most abstract level, one analyses what the system does from a computational point of view, or, more generally, what it appears to be trying to accomplish. Organisms are both products and continuing subjects of biological evolution, hence evolutionary considerations are likely to be important at this level. The second level concerns the way by which overall goals could be realized in terms of algorithms, or system-level procedures, of a kind that can work. Thus, this level translates the abstract level into attainable mechanisms, which perhaps approximate the ideal only roughly. Finally, the system needs to be realized materially, which is the level of implementation or realization. At this level, one expects a neurobiological model that can be directly compared with measurements.

All three levels will be addressed below, although only in broad outline. The theory, as it stands, should be seen primarily as a draft proposal rather than as a detailed blueprint. It aims to explain consciousness in its primordial form, that is, the form in which it presumably was first established in evolution and in which it presumably still exists today—close to the transition between non-conscious and conscious life forms, as well as close to the beginning of consciousness during the development of any conscious organism. Primordial consciousness is argued to be communicative rather than perceptual. Elaborate forms of consciousness, such as occur in adult humans, are not discussed here. Nevertheless, the theory offers a clear explanation of why consciousness is experienced and how that may be realized in the neurobiology of the brain. The neurobiological part of the theory shares characteristics with theories that are closely associated with neurobiological observations, such as those mentioned above (see further Section 10). Yet, the computational and algorithmic parts of the theory are novel, to my knowledge, and important details of the neurobiological proposal are as well.

Many current theories of consciousness are associated with philosophical analysis. The focus of this article is on computation and neurobiology, hence addressing philosophical theories is beyond its scope. Suffice it so say here that the two main possibilities considered by the latter, that consciousness is produced either by having representations of reality or by having representations of such representations, do not apply here. Although the first processing stage discussed below can be viewed, roughly, as producing representations (but of a special kind), the second stage—the one conjectured to produce consciousness—does not produce representations, but rather their inversion. What is meant by the latter should become clear below.

## 2 Preview of the explanation

The explanation below consists of a series of increasingly detailed theoretical elaborations, which may, at first reading, seem unconnected to consciousness and its neurobiology. I will therefore briefly state here where the argument leads to, and why the elaborations are needed. The argument concludes in Section 9 with the conjecture that consciousness is a transient and distinct cause that is produced when an individual prepares to communicate—externally or internally—a particular class of internal variables. Such preparation requires an operation that can be realized neurobiologically by a dual use of the thalamocortical feedback loop (Section 8). The particular class of communicated internal variables is rather special, because they estimate components of the evolutionary fitness of the individual itself. This is explained in Sections 3 and 4, and how to prepare these internal variables for communication is explained in Sections 5-7. The explicit dependence on fitness is crucial, because it is, probably, the only way by which real (literally present) estimation can arise and can be sustained in nature.

## 3 An internal estimate of evolutionary fitness

A major feature of any biological organism is its evolutionary fitness. Depending on the application, fitness is defined and used in various ways in biology (see, e.g., Endler 1986, p. 40, for examples). Often it is used as a purely statistical concept (as in population genetics), but alternatively it may be defined in a more mechanistic way, as a property of individual organisms. The latter is chosen here, where fitness is understood basically as an organism's propensity to survive and reproduce (quantifiable by combining expected lifetime and reproductive rate). More generally, it quantifies—as a statistical expectation—how well an organism may transfer its properties to other organisms, in particular to those of subsequent generations. This leads to generalizations such as inclusive fitness (which includes helping related organisms, such that related genes can get reproduced, see Section 5) and fitness effects produced by social and cultural transfer of properties. Importantly, fitness, as used here, is a forward-looking, probabilistic measure; actually realized survival and reproduction subsequently vary randomly around the expected value. Fitness is taken to change from moment to moment, depending on circumstances (e.g., the availability of food) and on the state of the organism (e.g., its health). Evolution by natural selection occurs when organisms in a population vary with respect to their typical fitness, on the assumption that at least part of that fitness is produced by heritable traits.

The precise form that fitness takes is not crucial for the theory below, as long as it is a reasonably adequate, forward-looking (i.e., predictive) measure of evolutionary success. Fitness $f$ is then assumed to be the outcome of a highly complex physical process, which can be written formally as an operator $F$ acting upon the relevant part $w$ of the world

$$f = Fw. \tag{1}$$

Here $f$ is a scalar function of time; $F$ and $w$ each depend on time as well, but their detailed form is left unspecified here—the fitness process is assumed to be nonlinear and nonstationary, and of high and time-varying dimensionality. The effective world $w$ encompasses not only external circumstances, but also the internal state and structure of the organism itself. The latter is called here the (biological) form of the organism. As an aid to the reader, Table 1 provides a list of the symbols used in this article, as well as a summary of their meaning.

When circumstances change, $f$ may change as well. If it decreases and such a decrease is indirectly detected by the organism (e.g., when food becomes scarce), then this usually engages compensating mechanisms (e.g., by switching to other food sources or by changing metabolic rates). Such compensating mechanisms can be viewed as forms of phenotypic plasticity (e.g., Nussey et al. 2007). Phenotypic plasticity refers to systematic changes of an organism's form during its lifetime, which includes, for example, changes in behavioural dispositions. Compensating mechanisms may either be fully inherited (when they originate from previous evolution) or not or partially inherited (such as when they are mostly established by previous learning by a particular organism). In either case, they respond to a problem that has occurred before, presumably many times. Inherited or learned compensating mechanisms are not further considered here but are merely acknowledged as an established baseline. The mechanism discussed below is taken to work on top of this baseline.

When circumstances change in an unexpected way, such that no ready-to-go compensating mechanisms are available or can be readily learned, the organism may still need to respond. Such a response can only be random and undirected. Yet, even if the proper direction of the response is not known, this is not true of the proper mean magnitude of the response. The following qualitative considerations make this plausible (for quantitative work see van Hateren 2015a–c). When $f$ becomes large as a result of changing circumstances, there is little reason to change an organism's form. It is already performing well, and even improving. On the other hand, when $f$ becomes small as a result of changing circumstances, not changing an organism's form may soon result in death. Then, it is better to change its form. Although this may initially lead to even lower $f$, it also increases the chances that a form with higher $f$ is found—perhaps after continued change. On average, this is still better than not changing at all and waiting for a likely death. Thus, the variability of changing an organism's form should be a decreasing function of $f$: large variability when $f$ is small ('desperate times call for desperate measures', if desperate includes undirected) and small variability when $f$ is large ('never change a winning team', or at least not much). Note that changes are made in a random direction, and that only the statistics of their magnitude (i.e., the variance) is modulated. This means that the mechanism acts in a slow, gradual way, not unlike the stochastic process of diffusion. The stochastic changes let the organism drift through an abstract, high-dimensional space of forms, drifting faster where fitness is low and slower where fitness is high. In effect, it lets the form of an individual organism move away from low-fitness forms (because variability is high there) and stay close to high-fitness forms (because variability is low there).

**Table 1** Terms used in the theory

| Symbol | Meaning | Comments |
|---|---|---|
| $f \in \mathbb{R}_{\geq 0}$ | Fitness | $f(t)$, varies in time |
| $w \in \mathbb{R}^K$ | Relevant part of the world | $w(t)$, varies in time;<br>$K$ is huge and varies in time |
| $F$ | Fitness operator | $f = Fw$, $F$ transforms $w(t)$ to $f(t)$;<br>$F$ is nonlinear and nonstationary |
| $f_i \in \mathbb{R}^{p_i}$ | A component of the fitness $f$ | $f_i(t)$, with $f(t)$ produced by transforming $\{f_i(t)\}$;<br>$p_i$ and the number of components vary in time |
| $w_i \in \mathbb{R}^{q_i}$ | A component of the world $w$ | $w_i(t)$, with $w(t)$ produced by transforming $\{w_i(t)\}$;<br>$q_i$ and the number of components vary in time |
| $F_i$ | A componentwise fitness operator | $f_i = F_i w_i$, $F_i$ transforms $w_i(t)$ to $f_i(t)$;<br>$F_i$ is nonlinear and nonstationary |
| $x \in \mathbb{R}_{\geq 0}$ | Individual's estimate of own fitness | $x(t)$, varies in time |
| $u \in \mathbb{R}^k$ | Sensed part of the world | $u(t)$, varies in time;<br>$k$ is large (yet $k \ll K$) and varies in time |
| $X$ | Fitness estimator (i.e., estimation operator) | $x = Xu$, $X$ transforms $u(t)$ to $x(t)$;<br>$X$ is nonlinear and nonstationary |
| $x_i \in \mathbb{R}^{n_i}$ | A component of the fitness estimate $x$ | $x_i(t)$, with $x(t)$ produced by transforming $\{x_i(t)\}$;<br>$n_i$ and the number of components vary in time |
| $u_i \in \mathbb{R}^{m_i}$ | A component of the sensed world $u$ | $u_i(t)$, with $u(t)$ produced by transforming $\{u_i(t)\}$;<br>$m_i$ and the number of components vary in time |
| $X_i$ | A componentwise fitness estimator | $x_i = X_i u_i$, $X_i$ transforms $u_i(t)$ to $x_i(t)$;<br>$X_i$ is nonlinear and nonstationary |
| $\bar{X}_i$ | Approximate inverse of fitness estimator $X_i$ | $\bar{X}_i \approx X_i^{-1}$ |
| $\hat{u}_i \in \mathbb{R}^{m_i}$ | Communicated report on $x_i$ | $\hat{u}_i = \bar{X}_i x_i$ |
| $u_i' \in \mathbb{R}^{m_i}$ | Report as received | $u_i' \approx \hat{u}_i$ |
| $X_i'$ | Receiver's version of $X_i$ | $X_i'$ and $X_i$ are similar |
| $x_i' \in \mathbb{R}^{n_i}$ | Receiver's estimate of sender's fitness component estimate | $x_i' = X_i' u_i' \approx x_i$ |
| $G$ | Gain operator | $G$ amplifies and transforms $(x_i - X_i \hat{u}_i)$ to $\hat{u}_i$ |
| $f^+ \in \mathbb{R}_{\geq 0}$ | Fitness-to-be | The increased fitness that gradually results from modulated stochastic changes |
| $Q_{X_i}$ | Quality of estimator $X_i$ | Accuracy by which $x_i$ estimates $f_i$ |
| $Q_{\bar{X}_i}$ | Quality of inversion of $X_i$ | Accuracy of $\bar{X}_i$ as an inverse of $X_i$ |

Although fitness is a feature of any organism, it is a factor that cannot be observed directly. The only way by which an organism can benefit from the above mechanism is when it contains an internal process that makes an estimate of its own fitness. Such an estimate is evolvable, because it increases subsequent fitness. Moreover, it is under evolutionary pressure to become and remain reasonably adequate as a predictor of evolutionary success. The estimate is called $x$ below, and the operator that produces it, $X$, is called an estimator. This corresponds to the modern, statistical use of the term: an estimator is a procedure (here $X$) that produces an estimate (here $x$) of a parameter (here $f$). It is the outcome of a complex physiological or neurobiological process, which can be written formally as the operator $X$ acting upon the part $u$ of the world that is accessible to the organism's sensors

$$x = Xu. \tag{2}$$

Here $x$ is a scalar function of time; $X$ and $u$ could be specified, in principle, in terms of physiological or neurobiological circuits and sensory inputs. $X$ is assumed to be nonlinear and nonstationary, and the dimensionality of $X$ and $u$ is assumed to be high and time-varying. Even though $X$ is complex, it is far less so than $F$, because the latter includes not only the physical environment, but also the organism itself, as well as other organisms around. Hence, $x$ can only approximately estimate $f$. Note that $x$ results from a distributed operator $X$, and is likely not localized, but distributed throughout the system. The effects of $x$ on the variability of the organism are likely distributed as well.

$X$ may be generally present in forms of life, as a physiological process. But for the present topic the relevant question is whether it could be a neurobiological process in those life forms that have an advanced nervous system. There are indeed strong indications that operations similar to $X$ may be present in the brain. Large sections of the brain are devoted to evaluations of biological value (Damasio and Carvalho 2013), and (predicted) reward can drive how neural circuits are adjusted (Glimcher 2011). Although these systems may work primarily according to already evolved and learned mechanisms, at least part of them may modulate variability (e.g., of behavioural dispositions) in the way explained above. Then, variability is expected to be a decreasing function of value (insofar that represents $x$). However, the topic of this article is not how $X$ is realized neurobiologically, but rather how $X$ can produce consciousness, once we take its existence as plausible or at least quite possible. Thus, below we will assume that $X$ and $x$ are present and see where that leads us.

## 4 Components of *X* and *F*

Equations (1) and (2) represent the overall processes that produce and estimate fitness, respectively. In practice, $X$ needs to be produced in manageable chunks. For example, part of fitness may depend on the ability to catch visually spotted prey, thus involving the visual system (detecting, discriminating, and tracking prey) and the motor system (chasing and capturing prey). Other parts will depend on other senses, or on evaluating the internal physiological state of the organism (e.g., states associated with thirst or pain). We will assume here that $X$ is parsed into a large set of such components, $X_i$, where the index $i$ denotes different ones. The components of $X$ are estimators that correspond to components $F_i$ of the $F$ operator. Formally, we can write

$$x_i = X_i u_i \tag{3}$$

and

$$f_i = F_i w_i, \tag{4}$$

where $u_i$ is a subset of $u$, and $w_i$ of $w$. Here $x_i$ estimates $f_i$, but because neither will be a scalar in general, this is a more complex version of estimation than above for $x$ and $f$. Importantly, the accuracy by which $x_i$ estimates $f_i$ is not directly relevant, but only insofar as it contributes to the accuracy by which $x$ estimates $f$. It is only the latter estimate that can gradually increase fitness through the stochastic mechanism explained in Section 3.

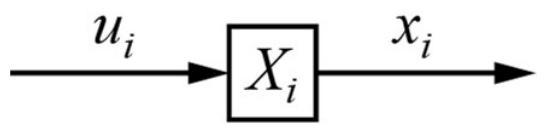

**Fig. 1** Diagram representing the equation $x_i = X_i u_i$, with $X_i$ an operator acting upon $u_i$ and producing $x_i$.

Below, I will make extensive use of diagrams for explaining the theory. Figure 1 shows the one corresponding to Eq. (3). In all diagrams, symbols in boxes denote operators, arrows denote the flow of processing, and symbols next to the arrows denote quantities on which the operators act (as input) or which the operators produce (as output). When a diagram would be used to model a specific system, it could be readily translated into a full quantitative model. However, that requires specific assumptions, such as on dimensionality, type of nonlinearity, and type of nonstationarity. Full quantitative modelling will not be attempted here, first, because it would be premature at this early stage of the theory, and, second, because the required level of neurobiological detail does not yet exist. As stated in Section 3, $X$ is assumed to be nonstationary, and this applies to each $X_i$ as well. Not only the specific form of an operator $X_i$ changes over time, but also the way by which it interacts with other parts of $X$. Part of these changes are regulated by the evolved and learned mechanisms mentioned before, but another part is produced by the variability that is controlled by $x$. The latter drives variations of the form of the organism, which includes variations of $X$, because $X$ is part of the organism. As a result, the form of $X_i$ can change in ways that are only partly predictable. In order to stress the nonstationarity of $X_i$, it will be written explicitly as a function of time in the diagrams below, as $X_i(t)$ (Fig. 2). Such nonstationary changes in $X_i$ will be slower than the fast changes as a function of time that one can expect in $w$, $f$, $w_i$, $f_i$, $u$, $x$, $u_i$ and $x_i$. The presence of fast changes is to be understood implicitly, but such time dependence will not be written explicitly in order to reserve $t$ for nonstationary changes.

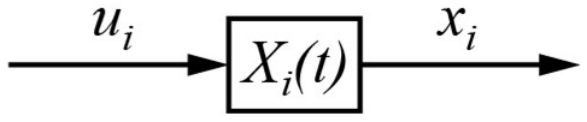

**Fig. 2** Explicitly writing $X_i(t)$ as a function of time $t$ is used to indicate that $X_i$ is nonstationary and changes slowly.

## 5 Enhancing fitness by communicating one's estimates

Fitness was described above, in its simplest form, as an organism's propensity to survive and reproduce. Although this may be valid for some species, fitness is often more complicated. A major extension of fitness occurs when organisms help closely related organisms. If the reproductive success of a helped organism increases as a result, this can indirectly increase the fitness of the helping organism. This is so, because the helping organism shares many genes with the offspring of the helped organism. Thus, the helping organism indirectly promotes disseminating its own properties. If this fitness benefit outweighs the cost of helping, then it is a worthwhile strategy from an evolutionary point of view. Fitness that includes this extension is known as inclusive fitness (Hamilton 1964). Inclusive fitness is still a property of each individual organism. It is to be taken, along with the benefits it can produce, in a statistical, probabilistic sense. Benefits need not always occur, but they are expected, on average. Below, fitness and $f$ refer to inclusive fitness, and $x$ is then an estimate of inclusive fitness. For the explanation of primordial consciousness, we will assume here the simple case of two closely related individuals that can mutually benefit from this mechanism, by cooperating with each other.

Cooperation often relies on communication between the cooperating individuals. Cooperative benefits then depend on exchanging useful messages, such as about the environment or about behavioural dispositions. One possibility for such communication is that it is hardwired or otherwise ingrained in the organism's physiology (e.g., through learning). Then communicative behaviour is either fixed or can only be learned within the narrow margins of fixed constraints. Moreover, the information that is transferred is then fixed as well, and it does not involve an $X$-process that drives random, undirected changes of an organism's form. Examples of this type of hardwired communication are quorum sensing in bacteria and the food-pointing waggle dances of honey bees. Such stereotyped transfer of information is not further discussed here.

We will assume here that all non-stereotyped communication is based on communicating factors $x_i$ (like the one in Fig. 2). When such communication is performed in a cooperative setting, the fitness of both sender and receiver is likely to benefit, as is argued now. A factor $x_i$ estimates a corresponding factor $F_i w_i$. The latter depends on the state of the world (via $w_i$) as well as on how this affects the sender (via $F_i$). Communicating an estimate of this (i.e., $x_i$) is likely to enhance fitness for two reasons. First, it may directly increase $f$, in a similar way as a stereotyped transfer of useful information can increase $f$. Second, it is likely to increase the accuracy by which the receiver can estimate, through its own process $X$, those parts of the sender's $X$-process that are relevant to the cooperation (parts to which $x_i$ belongs). As a result, the receiver's $x$ becomes more accurate as an estimate of $f$, because parts of $F$ are determined by the cooperation. A more accurate $x$ subsequently and gradually increases the receiver's fitness through the mechanism explained in Section 3. The fitness of the sender is then likely to increase as well, because of the cooperation. This mutual effect is further enhanced when sender and receiver engage in a dialogue, as will typically happen (where 'dialogue' is taken here and below as nonverbal, because of the focus on primordial consciousness). We conclude that $X$-based communication is likely to enhance fitness. Hence, it is evolvable and assumed here to be present. Examples of cooperative settings that support this type of communication in its most primordial form are mother-infant bonds in mammals and pair bonds in breeding birds. On average, fitness increases either directly (as for infants) or indirectly (because offspring is supported).

Importantly, $x_i$ is an internal factor of the sender. From here on, we will assume that it is internal to the sender's brain. The reason for this assumption is that the theory that is explained below requires quite complex transformations (such as inversion of operators). These transformations are at a level of complexity that is presumably only realizable in advanced nervous systems (and not in multicellular organisms without a nervous system, nor through processing within unicellular organisms). A sender communicating its $x_i$ to a receiver requires that the receiver obtains access to a factor that is, in effect, similar to $x_i$. The question is, then, how the sender can communicate $x_i$ in such a way that it will produce a similar factor, called $x_i'$ below, in the brain of the receiver. Specific hardwired solutions, such as in the case of honey bee dancing, are not viable here, for two reasons. First, $X_i$ is nonstationary and thus cannot be anticipated in detail, and, second, the number of factors $x_i$ that are potential candidates for communication may be huge.

A viable way to communicate $x_i$, at least approximately, is the following. It is reasonable to assume that the receiver has an operator $X_i'$ that is similar to the one of the sender, $X_i$. This assumption is reasonable, because the cooperative setting presupposes that the two individuals are similar, that they share similar circumstances and possibly a similar past, and that they share goals to which the communication is instrumental. As is shown now, the sender can then communicate $x_i$ by utilizing an operator $\bar{X}_i$ that is approximately the inverse of $X_i$, thus

$$\bar{X}_i \approx X_i^{-1}. \tag{5}$$

Recalling that $x_i = X_i u_i$ (Eq. 3) this implies that (see the left part of Fig. 3)

$$\hat{u}_i = \bar{X}_i x_i \tag{6}$$

and

$$\hat{u}_i \approx u_i. \tag{7}$$

Subsequently, the sender communicates $\hat{u}_i$, which is possible because it belongs to the same space as $u_i$, that is, the space accessible to sensors and motor outputs. Assume that the receiver gets $u_i' \approx \hat{u}_i$, and subsequently applies $X_i'$ to $u_i'$. Then this results in

$$x_i' = X_i' u_i' \approx X_i' \hat{u}_i \approx X_i \hat{u}_i \approx X_i u_i = x_i, \tag{8}$$

or

$$x_i' \approx x_i. \tag{9}$$

Figure 3 shows the diagram that corresponds to this way of communicating.

Several concerns should be mentioned here. In this simple case, one might object that the receiver would be better off by directly observing $u_i$ rather than the communicated $\hat{u}_i$. However, below we will

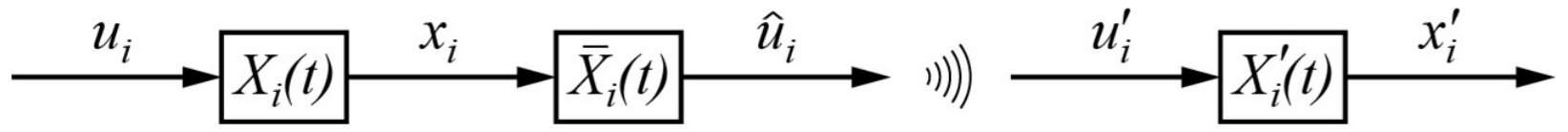

**Fig. 3** A sender (left) communicates $x_i$ by utilizing an approximate inverse $\bar{X}_i$ of its $X_i$ operator, with the receiver (right) approximately reconstructing $x_i$ by applying its own $X_i'$ to consider elaborations of this basic diagram for which that would not work. A second concern is whether the sender would be able to reproduce $u_i$ accurately, with the means of expression available. We will assume here simple cases where that is possible. For example, if $u_i$ is a facial expression (such as a smile) observed in another individual, then this could be mimicked (e.g., reciprocally by mother and infant). The same goes for sounds, touches, and expressions of emotions (such as laughter and crying) that are produced by other individuals. However, if $u_i$ is part of the physical world, such as a general visual scene, $\hat{u}_i$ may have to be produced by indirect means, using learned conventions that are understood by both sender and receiver. This applies also to other complex communications, for instance when social situations or language are involved. How such conventions can be gradually learned is beyond the scope of this article, because we focus here on primordial consciousness (but see van Hateren 2015d for elaborations and references).

A final concern is whether $X_i$ has an inverse at all. In general, this is not guaranteed. If an exact inverse does not exist, then it is often possible to define an operator that at least minimizes the distance between $u_i$ and $\hat{u}_i$ (analogous to the pseudoinverse of matrix operators). However, the more serious problem is not whether there is a mathematical solution, but rather whether there is a plausible and realistic neurobiological mechanism that could invert $X_i(t)$. This is a major issue, because $X_i(t)$ is nonstationary and not known in advance. Yet, a possible solution is proposed next.

## 6 An algorithm for inverting the unknown

The presence of $X(t)$ and its components $X_i(t)$, and the capability to utilize them, are evolved properties. But what their specific form will be at any point in time is unknown, even if $X(t)$ and $X_i(t)$ may be subject to broad constraints. The key point of $X(t)$ and $X_i(t)$ is that they are highly adaptable during an individual's lifetime. How, then, could an inverse $\bar{X}_i(t)$ evolve, or even a procedure to learn such an inverse, if it is not well defined of what exactly it should be the inverse? The capacity to produce a highly adaptable inverse seems to require the capacity to track the detailed structure of $X_i(t)$, which would be extraordinarily hard to evolve (given the nonstationarity of $X_i(t)$). Fortunately, there is a much easier way to invert $X_i(t)$, which does not even require a separate $\bar{X}_i(t)$, and which seems readily evolvable. It is based on an old electronics trick that uses feedback for inverting an operator.

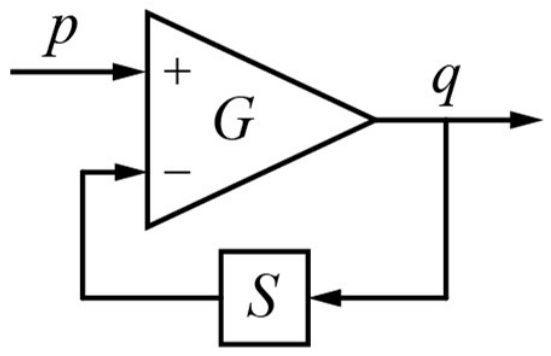

**Fig. 4** Circuit that inverts an operator $S$ by using it in the feedback path to the inverting input of an operational amplifier with high gain $G$.

Figure 4 shows the basic idea. The triangle symbolizes an operational amplifier with high gain $G$. The device amplifies the difference between the voltages on its plus and minus inputs. Suppose $S$ is an operator that transforms $q$ (which is output of the amplifier as well as input of $S$) and feeds the result back to the minus input of the amplifier. If the input is $p$, then one finds

$$G(p - Sq) = q, \tag{10}$$

hence

$$p = Sq + q/G. \tag{11}$$

If the gain $G$ is sufficiently large such that $q/G$ is small compared with $Sq$, then

$$p \approx Sq \tag{12}$$

thus

$$S^{-1}p \approx S^{-1}Sq = q. \tag{13}$$

This depends on the assumption that $S^{-1}$ exists, such that $S^{-1}S$ is the identity operation (or close to the identity operation if $S^{-1}$ can only be approximated). Equation (13) shows that the output $q$ of the circuit in Fig. 4 approximately equals $S^{-1}$ acting upon the input $p$. In other words, the circuit as a whole operates, approximately, as the inverse of the $S$ operator. Apart from the conditions that the gain is sufficiently large and that the inverse of $S$ exists, there are no further conditions on the nature of $S$: it could be an arbitrary nonlinear and nonstationary operator. A general discussion of operators, including inversion of the type shown here, can be found in Zames (1960, p. 19).

Although the circuit of Fig. 4 has a scalar input and output (because a voltage is a scalar), we assume here that it can be generalized to higher dimensions, including different dimensions for input and output. For the present purpose, $X_i$ must play the role of $S$ (because the approximate inverse $\bar{X}_i$ is desired), $x_i$ that of the input $p$, and $\hat{u}_i$ that of the output $q$. According to Table 1, the dimensionality of input and output is then $n_i$ and $m_i$, respectively, which are both expected to be large. Thus, the operators $G$ and $X_i$ transform between high-dimensional spaces. One consequence of the assumed high dimensionality of the operators is that the neurobiological interpretation given in Section 8 below should not be regarded purely at the single neuron level. Rather, the operators are realized by the action of groups of interacting neurons, and the quantities these operators act upon and produce are compound signals of groups of neurons as well.

The requirement that $S$ has an inverse—or at least an approximate inverse—puts constraints on the types of processing allowed. Moreover, if the inverse is produced by feedback, as is proposed here, time delays can produce problems. Delays are inevitable when signals are transferred across the brain. As a result, the feedback may become unstable, unless the speed of processing by $S$ is significantly slower than the delays. Processing by $S$ is, thus, likely to be slow, for this reason alone (if not for other reasons as well, such as the slowness and serial nature of communication).

The circuit of Fig. 4 is a minimalistic one, primarily intended for explaining the basic idea. It is quite conceivable and likely that it could be elaborated to specifically suit the strengths and limitations of the neurobiological circuits that realize $X_i$. Thus, Fig. 4 should be seen as a first-order draft, rather than as a finished end product. Nevertheless, it is a suitable basis for developing the theory further.

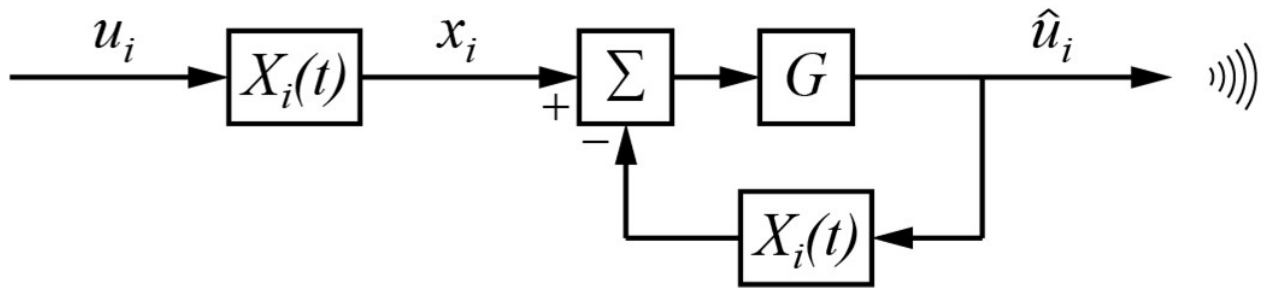

**Fig. 5** The right part of the circuit uses the mechanism of Fig. 4 for producing an approximate inverse of $X_i(t)$.

Figure 5 shows how this mechanism can be applied to the sender's part of Fig. 3. The amplifier of Fig. 4 has been split here into two operations, a summation (Σ) of one input with a plus sign and the other with a minus sign, followed by a gain $G$. As can be seen, $X_i(t)$ is used twice, once in the regular forward direction, and once in the return path of a feedback loop in order to produce $\bar{X}_i(t)$. However, the circuit of Fig. 5 is not quite right, because a single neurobiological instance of $X_i(t)$ cannot perform both operations simultaneously. The circuit should be modified such that $X_i(t)$ switches between its two roles. Figure 6 adds switches that allow $X_i(t)$ to be utilized in either of two states: state 1 for the forward direction, and state 2 for the inverse. The circuit works by continually switching between these states. Because of the switched processing, buffers are needed in order to retain the results of the previous state. Consequently, the buffered versions of $x_i$ and $\hat{u}_i$ have to follow a slower time course than their inputs $x_i$ and $\hat{u}_i$. This is indicated by writing the buffered versions explicitly as a function of time, in a similar way as was introduced above for $X_i(t)$. A side effect of switching is that the forward $X_i(t)$ and the one producing the inverse $\bar{X}_i(t)$ are not evaluated at exactly the same time, and thus cannot produce an exact inverse. However, the resulting error will be small if $X_i(t)$ changes only slowly compared with the switching rate (which may be in the order of 10 Hz in the primate brain, see, e.g., Koch 2004, pp. 264–268).

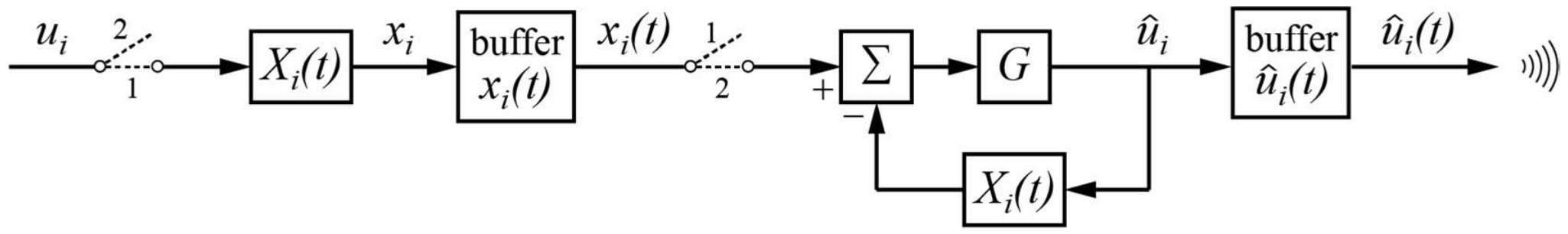

**Fig. 6** The circuit of Fig. 5 can work with the same instance of $X_i$ if it switches continually between state 1 (forward operation of $X_i$) and state 2 (inverse operation of $X_i$). Buffers are needed for retaining the result of the previous state.

## 7 From monologue to dialogue and internal dialogue

Figure 6 showed how the sender can produce a monologue directed at a receiver. The receiver may then respond by using the same type of processing. This can lead to a continued dialogue as pictured in Fig. 7. The upper path shows the sender, and the lower one the receiver. It assumes that the dialogue is in full swing, having been started at an earlier point in time. The dialogue progresses, because both $X_i(t)$ and $X_i'(t)$ are nonstationary, and, in addition, change in ways that are not fully identical in the two individuals. This also applies to the buffers of internal estimates ($x_i(t)$ and $x_i'(t)$) and of communicated signals ($\hat{u}_i(t)$ and $\hat{u}_i'(t)$).

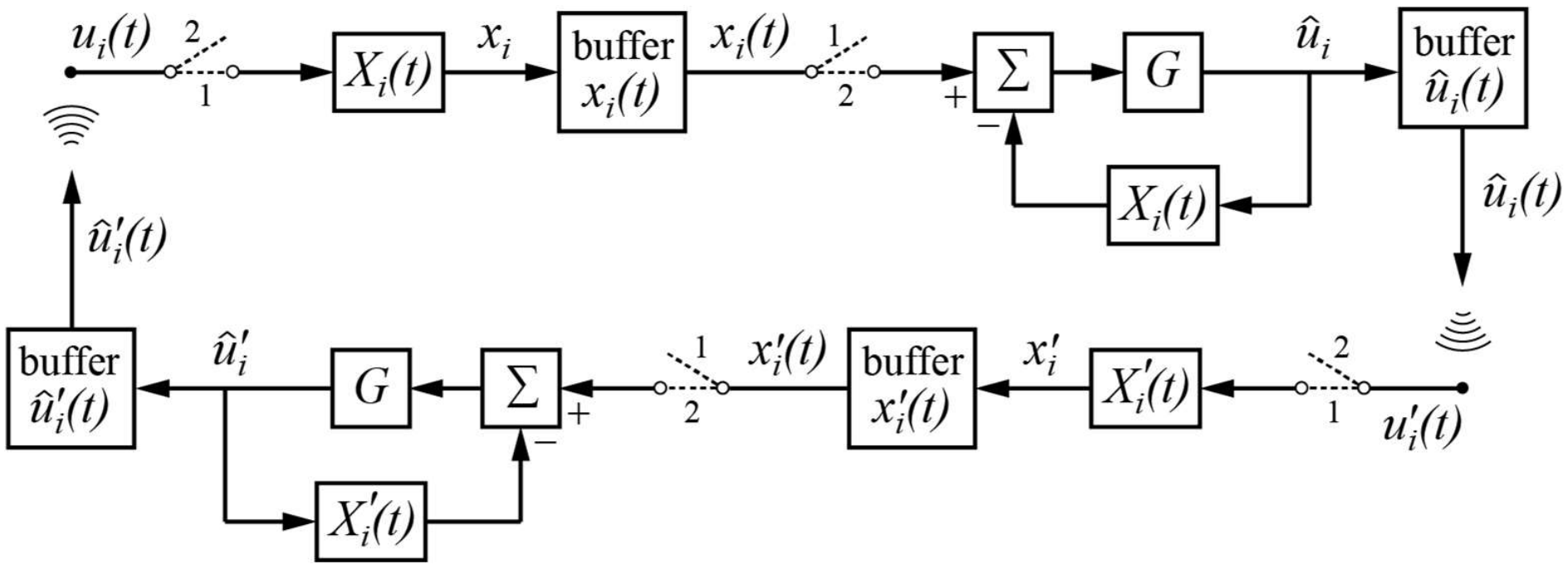

**Fig. 7** Dialogue between a sender (upper path) and a receiver (lower path).

Communication could be improved if the sender would not invert its own $X_i(t)$, but a prediction of the $X_i'(t)$ that the receiver is likely to use when receiving the communication (and then the receiver should do something similar in return). This requires that the communicating partners maintain a model of each other's $X_i$, and that part of their communication is used to keep these models up to date. Elaborations along these lines will not be further discussed here (but see van Hateren 2015d for draft proposals and references to similar ideas).

Rather than engaging in a dialogue with a communicative partner, an individual may engage in an internal dialogue. This is produced when the output of the upper path of Fig. 7 folds back to provide its own input (Fig. 8). The model of Fig. 8 somewhat resembles a two-cycle (two-stroke) combustion engine, as it cycles through filling the buffer $x_i(t)$ at stage 1, then producing $\hat{u}_i(t)$ at stage 2 ('ignition'), then a renewed filling at another stage 1, and so on. Because $X_i(t)$ gradually changes over time, $x_i(t)$ and $\hat{u}_i(t)$ do so as well. Here $x_i(t)$ is a strictly internal variable of the brain, even if it estimates an external part of reality (namely $F_i w_i$). On the other hand, $\hat{u}_i(t)$ does not estimate anything external, but just belongs to the individual's sensorimotor space that can be observed or acted upon. This space is shared with conspecifics. Hence, $\hat{u}_i(t)$ can be communicated, and is, thus, instrumental to communicate $x_i(t)$. The latter, being an estimate, does not belong to sensorimotor space, and could not be communicated directly.

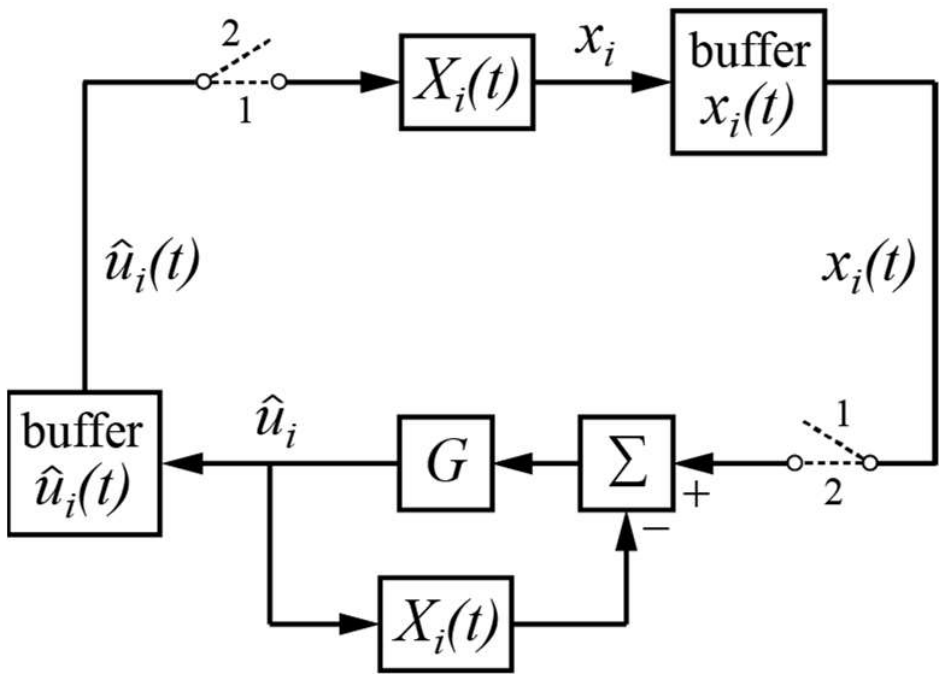

**Fig. 8** Internal dialogue, switching between forward and inverse uses of $X_i$. The results of the internal dialogue, $x_i$ and $\hat{u}_i$, change through time because $X_i$ is nonstationary.

Evolving a purely stand-alone version of Fig. 8 is unlikely from an evolutionary point of view, because there would be no benefits obtained from producing $\hat{u}_i(t)$. Any processing that might increase fitness could be performed at the level of $X_i(t)$ and $X$, without going to the trouble of inverting $X_i(t)$. Hence, the mechanism of Fig. 8 is only useful if it is, at some point at least, combined with the dialogue of Fig. 7. For example, cycling through the loop of Fig. 8 may prepare $X_i$, and consequently $X$, for more adequate future interactions according to Fig. 7. More adequate means here having a better chance of increasing fitness, at least on average. The loop of Fig. 8 can also explain why not only senders are conscious, but receivers as well, and why non-communicative stimuli (such as a general visual scene) can be consciously perceived. In both cases, the stimuli may induce cycling through the loop, and thus produce consciousness—again, with the ultimate prospect of communication and dialogue with a partner.

## 8 Neurobiological interpretation: A two-cycle thalamocortical loop

If the theory is correct, then it should be possible to identify its parts in the brain (I will be focussing here on the mammalian brain; see Section 10 for some comparative considerations). In other words, what could be a plausible neurobiological realization of the theory? I will discuss here what I view to be the most likely one, in terms that should be amenable to direct empirical tests (see Section 10). My interpretation is much informed by other neurobiological theories of consciousness, but making these connections will be postponed to Section 10.

A crucial part of the theory is that it contains a massive feedback loop, namely the one that inverts $X_i$ (as in Figs. 5-8). This loop is massive, because it applies to all components $X_i$ that together constitute $X$. The number of such components must be huge, pertaining to any part of the brain that could be important for estimating fitness. There are, of course, many feedback loops in any part of the brain, but two of the most conspicuous ones are the loop from thalamus to cortex and back, and the loop from cortex to striatum (and other parts of the basal ganglia) to thalamus and back to cortex. The latter loop is particularly involved with learning, reward, value, and adjusting behavioural dispositions (Chakravarthy et al. 2010; Hikosaka et al. 2014). It seems particularly well suited, then, to implement at least part of the (nonstationary) $X(t)$ operator. Thus, I will focus here on the thalamocortical loop as a candidate for the proposed feedback loop of Figs. 5-8.

The basic structure of the thalamocortical loop is sketched in Fig. 9a (Llinas et al. 1998; Butler 2008; Ward 2011; Sherman 2016). The dorsal thalamus contains nuclei that relay sensory inputs (in particular visual, auditory, and somatosensory ones) to the cortex, as well as a range of nuclei that receive cortical inputs that are relayed back to cortex. Connections from thalamus to cortex are marked by TC in Fig. 9a, where grey disks symbolize neurons or groups of neurons, and arrow heads symbolize excitatory input. In addition, there is massive feedback from cortical areas to corresponding parts of the thalamus, either directly to the dorsal nuclei (connection CT in Fig. 9a) or indirectly via the Thalamic Reticular Nucleus (TRN; connection CR). Neurons of the TRN provide inhibitory inputs to thalamic relay neurons (marked with a minus sign in Fig. 9a; connection RT).

This circuit seems to be well suited to produce the two-cycle mechanism explained above. In the first stage either sensory or cortical input is transferred to the cortex, where it engages $X_i$ and $X$ via intracortical connections and the cortico-striato-thalamo-cortical loop (including inputs from other nuclei, such as those in the upper brainstem). The result $x_i$ may be retained in a buffer for a short time, perhaps of the kind usually ascribed to iconic memory (e.g., Koch 2004, pp. 201–203). The second stage only occurs if and when $x_i$ is prepared for communication. Then the buffered result is transferred back to the thalamus (via CT), and is subsequently sent for a second time to the cortex to engage $X_i$ and $X$ in a similar way as before. But now the result is subtracted, through the inhibitory action of the TRN (via RT), from the buffered earlier result. The difference is amplified and transferred to the cortex. The second stage effectively inverts $X_i$, and thus transforms the buffered $x_i$ into a signal that is in the same sensorimotor space as the original sensory or cortical input to the thalamus (lower part of Fig. 9a). Hence, it is suitable for communication, or for using once more as input to the thalamus. It should be noted that the above description is only a primordial sketch: a more elaborate model should incorporate the different functional roles of the various thalamic nuclei and how they interact with different cortical and subcortical areas. Moreover, the presence and form of the proposed mechanism may well vary across and within nuclei.

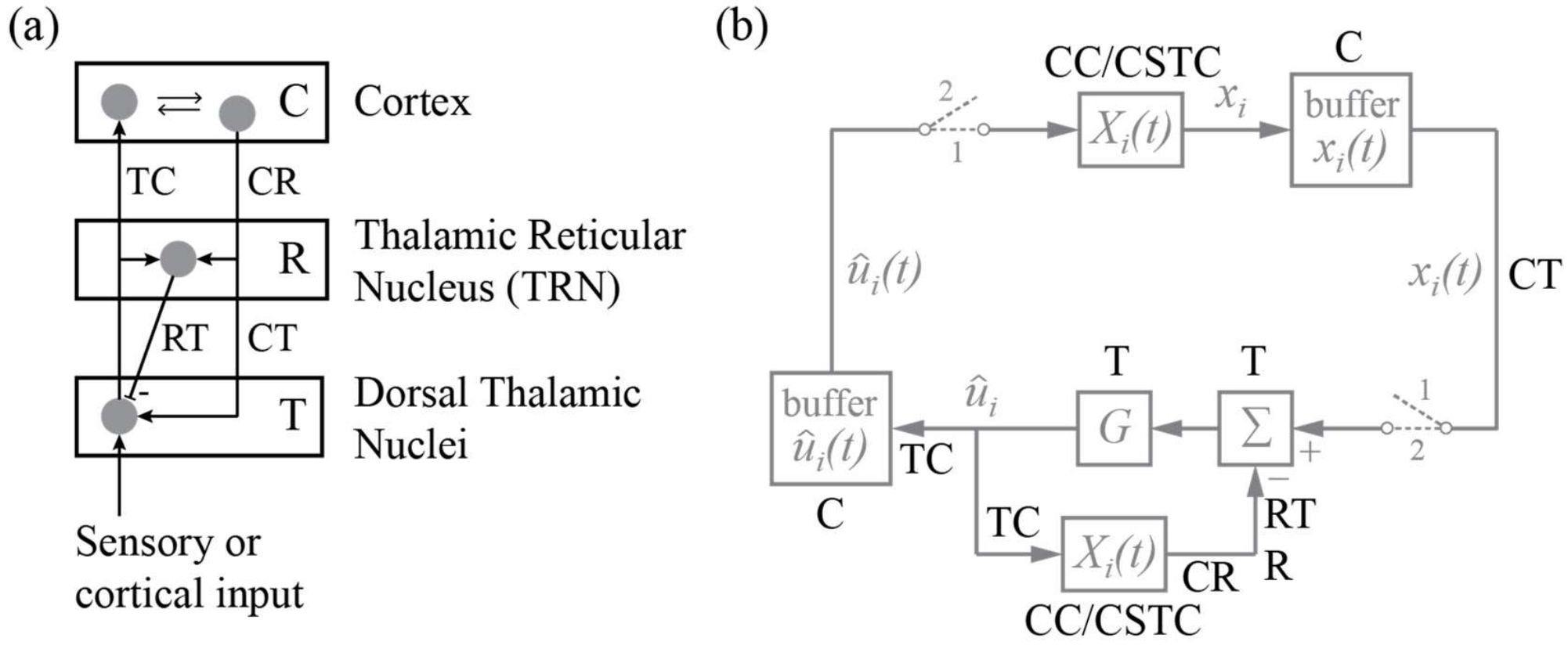

**Fig. 9** Neurobiological interpretation. **a** Basic structure of the thalamocortical feedback loop. **b** Tentative identification of the components of the model of Fig. 8 with the anatomical parts of **a**. See the main text for explanation.

Figure 9b tentatively identifies the components of the diagram of Fig. 8 with components of the thalamocortical circuit of Fig. 9a. It is not the only possibility for each and every component, and I will mention a few alternative configurations after I have explained the configuration as shown. $X_i$ is located in the cortex, produced through intracortical interactions (CC, engaging other $X_i$ as well) and, presumable, the cortico-striato-thalamo-cortical loop (CSTC). The buffers of $x_i(t)$ and $\hat{u}_i(t)$ are assumed to be located in the cortex. The upper left switch must then be in the cortex as well, but the lower right one could either be in the cortex or in the thalamus (as drawn). Subtraction and gain $G$ are shown to be located in the thalamus, but $G$ might as well be located in the cortex, or partly so. A complication of the diagram is that the CT connection at the right (output of the first stage) needs to be functionally separate from the CR/RT connection in the feedback loop at the bottom (performing the second stage). Thus, either there are at least two types of CT neurons performing these functions, or a single type uses a special mechanism to perform both functions. Alternatively, the $x_i(t)$ buffer might be located in the thalamus, which implies, in that case, that the function of a CT neuron would switch between stages. It would act to fill the buffer at stage 1 and would function in the feedback loop at stage 2. A possible issue is the linear operator (Σ), because GABAergic inhibition by TRN neurons may act more like a modulation (i.e., a gain control) than like a linear subtraction. However, gain control and subtraction can readily be made to lie on a continuous scale by having expansive and compressive nonlinearities in input and output.

The reader may note that the direct excitatory input from thalamus to TRN (left part of Fig. 9a) has not been given a function in Fig. 9b. One possibility is that the negative feedback it provides to the thalamus compensates for the gain $G$ when that is not needed in stage 1. However, there are many other local circuits in thalamus, TRN, and cortex that are not represented in Fig. 9b (Traub et al. 2005; Izhikevich and Edelman 2008). Speculating on the role of these circuits and on the role of the various cortical layers seems rather premature at this stage. The viability of the theory of Section 7 and the general idea of Fig. 9b should be tested empirically first. Several suggestions for such tests are given in Section 10.

## 9 The distinct cause that feels like consciousness

So far, we have considered a computational model (Fig. 3), an algorithm (Figs. 6-8), and a neurobiological realization (Fig. 9) of how $x_i$ could be communicated. But why would any of this produce consciousness? I will now argue that the model gives rise to two different types of cause that are each rather special. Subsequently, I will argue that only the second type of cause, the one associated with $\bar{X}_i$, is felt as consciousness. A 'cause' is taken here as a factor (such as a variable or a material process) that, when perturbed, typically produces systematic (possibly probabilistic) changes in another such factor (see, e.g., Pearl 2009). The latter factor is then called the 'effect' of the cause. For example, in Eq. (1), both the operator $F$ and the effective world $w$ can be viewed as causing (i.e., producing) the fitness $f$, because perturbing $F$ or $w$ affects $f$ in a systematic way.

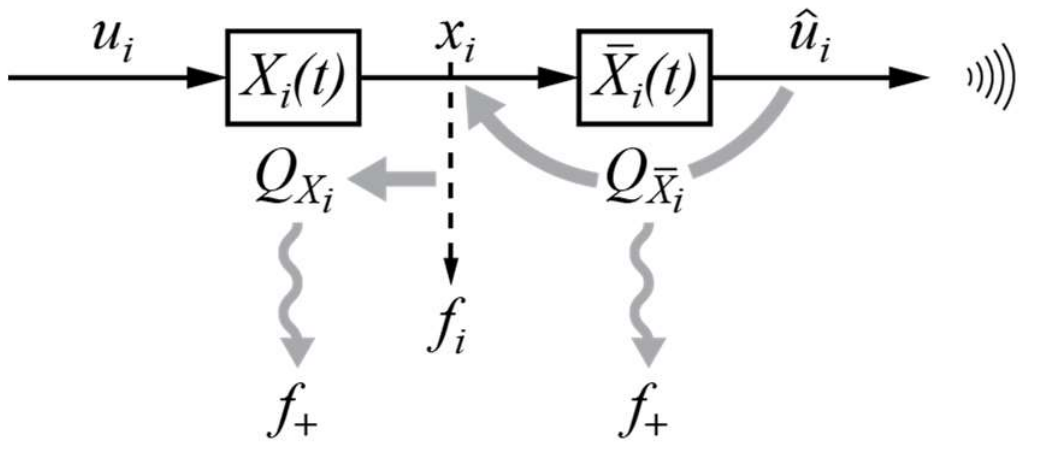

**Fig. 10** The sender's part of Fig. 3 contains two distinct causes associated with $X_i$ and $\bar{X}_i$, of which the second is conjectured to produce the feeling of consciousness. See the main text for explanation.

The sender's part of Fig. 3 is redrawn in Fig. 10, with a few additions. The dashed arrow denotes the fact that $x_i$ estimates $f_i$. Ultimately, this depends on the fact that $x$ estimates $f$. As explained in Section 3, the latter estimate causes a gradual increase of fitness through a stochastic mechanism. The increased fitness that gradually results is denoted below by 'fitness-to-be' (abbreviated to $f_+$, see also Fig. 10), whereas current fitness is denoted by $f$, as before. How effective this mechanism can be must depend on how well $x$ estimates $f$. When estimation is absent or poor, the mechanism cannot work or cannot work well. The better the estimator is, the higher $f_+$ will likely become. A simple way to quantify the goodness of an estimator is to evaluate the expected estimation error $\varepsilon_X$, that is, the expected difference between estimate ($x$) and estimated ($f$)

$$\varepsilon_X = E[\|x - f\|]. \quad (14)$$

Here $E$ denotes the expectation (expected value), and $\|\cdots\|$ denotes a norm, for example by taking the absolute value or by squaring. The goodness of the estimator is then inversely related to the error $\varepsilon_X$. In order to keep the presentation below as simple as possible we will avoid this inverse relationship by defining the quality $Q_X$ of the estimator $X$ as the reciprocal of the estimation error $\varepsilon_X$

$$Q_X = 1/E[\|x - f\|]. \quad (15)$$

The higher $Q_X$ is, the better the estimator is, and the higher $f_+$ can become eventually. This means that $Q_X$ should be regarded as a cause in the sense defined above: when $Q_X$ is perturbed (for example when $X$ learns to incorporate the fitness effects of a new part of the environment), this also changes $f_+$, at least on average. Moreover, an investigator may perturb and manipulate $Q_X$ on purpose (either empirically or in a model) and will then observe that $f_+$ is affected accordingly (in a systematic and statistically reliable way). Thus, $Q_X$ is a cause of $f_+$ (though, of course, not the only one). Because $Q_X$ directly depends on $x$, perturbing $x$ also affects $f_+$, as does perturbing $X$ (if this is assumed to be done in a way that changes $x$). Therefore, stating that $Q_X$, $x$, or $X$ is a cause of $f_+$ will all be regarded below as equally valid.

$X$ is a rather special cause, even though it has the characteristics of a regular neurobiological process. Such a process can be understood from how its parts work and how they interact. Thus, how $X$ works is just the compound result of how its material parts and their interactions work. However, the effect of $X$ on $f_+$ is *not* just the compound result of the effects of these parts ('effects of parts' is taken here and below to include interactions between parts). The reason is that the effect of $X$ on $f_+$ critically depends on the fact that $x$ is an estimate of $f$. If this estimation would be absent, there would be no effect on $f_+$ at all. Estimation is not fully defined by the material parts of $X$. In addition, it depends on what it estimates, fitness. One might then think that the effect of $X$ on $f_+$ would be just the compound result of the effects of the parts of the combined $X$ and $F$ processes. But this is not true either, which is explained next.

The effect of $X$ on $f_+$ is produced in an unusual, stochastic way. It depends on changing the form of an organism by modulating the variance of random, undirected change. As a result, it tightly couples a determinate factor, $x$, with an indeterminate factor (i.e., noise produced in an indeterminate way). Because this happens in a continually operating feedback loop (which modulates $x$ depending on earlier changes), the result is an inseparable mixture of determinate and indeterminate factors (van Hateren 2015a). The effect on $f_+$ is therefore not solely given by the combined $X$ and $F$ processes, but also by the effects produced by (modulated) noise (which may arise from thermal or quantum indeterminacy, or from intractable external disturbances). Importantly, the noise does not average out, but strongly determines the ultimate result. Moreover, the effect of $X$ on $f_+$ is not almost instantaneous (as it would be in any standard material process when considered at a microscopic scale) but arises only slowly and gradually by accumulating modulated stochastic changes of the brain. Hence, cause and effect occur on a stretched timescale, much longer than the timescale on which standard material processes work internally. Thus, $X$ is a rather exceptional cause, for several (related) reasons: it crucially depends on noise, it produces its effect quite slowly, and it depends on an evolved and sustained form of estimation.

In addition to the quality $Q_X$ of the overall estimator $X$, we can denote the quality of each estimator $X_i$ by an analogous factor $Q_{X_i}$. Because neither $x_i$ nor $f_i$ are scalars, there is no simple equivalent of Eq. (15). Nevertheless, it is clear that when $X_i$ changes and then produces an $x_i$ that estimates $f_i$ better (i.e., when $Q_{X_i}$ increases), that this will increase the overall $Q_X$ and, thus, will increase $f_+$. Therefore, each $X_i$ operator is a cause of $f_+$ as well, as is $x_i$. The effect of each $X_i$ depends on how $X_i$ contributes to the effect of $X$ on $f_+$. Because the latter effect is not just the compound result of the effects of the combined $X$ and $F$ processes, this is true of $X_i$ as well. This means that $X_i$ is a cause with an effect that is not just the compound result of the effects of its parts (i.e., of the combined $X$ and $F$ processes or any subset of those). The fat grey arrow on the left in Fig. 10 symbolizes the fact that the effect of $X_i$ (through its property $Q_{X_i}$) derives from how well $x_i$ estimates $f_i$. The downward wavy arrow symbolizes that it has a (slow, stochastically produced) causal effect on fitness-to-be ($f_+$).

Having established that $X_i$ is a cause with an effect that is not just the compound result of the effects of its parts, we can now investigate the consequences. We need to invoke here the basic notion that if something is a cause (in the sense that it is capable of producing material effects in the world), that one can then be certain that it exists in a real, literal way (i.e., other than merely as an abstract theoretical construct). However, most causes in nature have effects that are just the compound result of the effects of their parts. Such causes are, then, not distinct (i.e., not autonomous), because they do not produce anything beyond what their parts do (again, including interactions of parts). In contrast, $X_i$ produces an effect that is partly produced by modulated noise, and thus goes beyond what its parts (or the parts of $X$ and $F$ combined) produce. Hence, the cause $X_i$ (or, equivalently, the cause $x_i$) corresponds to something that not only exists literally, but also exists in a distinct way (i.e., as partly autonomous, because it adds something over and above the effects of its parts, due to the effects attributable to noise). Remarkably, it is not the only distinct cause in Fig. 10. There is a second one, marked by the curved grey arrow on the right, which is explained next.

The operator $\bar{X}_i$ produces $\hat{u}_i$ with the specific aim to communicate it, such that it can produce an $x_i'$ in the receiver that is similar to the $x_i$ of the sender (Fig. 3). The quality of $\hat{u}_i$ then depends on how well $\bar{X}_i$ can invert $X_i$. A perfect inversion would imply that $X_i\hat{u}_i$ would be equal to $x_i$. Therefore, the error $\varepsilon_{\bar{X}_i}$ made by operator $\bar{X}_i$ can be written as

$$\varepsilon_{\bar{X}_i} = E[\|x_i - X_i\hat{u}_i\|]. \tag{16}$$

Here $\|\cdots\|$ denotes again a norm, now not of scalars, but of the more complex quantities $x_i$ and $X_i\hat{u}_i$ (which are of the same type). We can now define the quality $Q_{\bar{X}_i}$ of the operator $\bar{X}_i$ as the reciprocal of the error $\varepsilon_{\bar{X}_i}$

$$Q_{\bar{X}_i} = 1/E[\|x_i - X_i\hat{u}_i\|]. \tag{17}$$

The higher $Q_{\bar{X}_i}$ is, the better $\bar{X}_i$ is, and the better $\hat{u}_i$ will serve the effectiveness of communication. A better effectiveness of communication is likely to increase the (inclusive) fitness of the sender, at least on average (and assuming that certain conditions apply, see Section 5). Again, the expected fitness effect of increasing $Q_{\bar{X}_i}$ is not immediate, because it depends on two individuals cooperating by utilizing the stochastic benefits of the communicated version of $x_i$ (either now, as in Fig. 7, or later, as is implicit in Fig. 8). Therefore, $Q_{\bar{X}_i}$ is a cause of $f_+$, which is indicated by the downward wavy arrow on the right of Fig. 10. Similarly as is argued above for $Q_X$, we may equally well say that $\bar{X}_i$ is a cause of $f_+$.

$\bar{X}_i$ has the characteristics of a regular neurobiological process, but, again, it is special. The only reason why $\bar{X}_i$ has an effect on $f_+$ is because it produces $\hat{u}_i$, and the only reason why producing $\hat{u}_i$ has an effect is because $\hat{u}_i$ can communicate $x_i$ (in its capacity to estimate $f_i$). Above, it was established that $x_i$ corresponds to something that exists in a distinct way. Thus, the effect of $\bar{X}_i$ on $f_+$ crucially depends on communicating something distinct (i.e., $x_i$), which implies that $\bar{X}_i$ is a cause with an effect that is not just the compound result of the effects of its neurobiological parts. Hence, the cause $\bar{X}_i$ corresponds to something distinct as well, but different from what corresponds to $x_i$ and $X_i$. The curved grey arrow on the right in Fig. 10 symbolizes the fact that the effect of $\bar{X}_i$ depends on how well $\hat{u}_i$ allows reconstructing $x_i$, and the downward wavy arrow indicates that the effect is on fitness-to-be, $f_+$.

We have now established that there are two distinct causes associated with the diagram of Fig. 10. However, these two causes have quite different spatial characteristics, as is explained next. The cause $X_i$ (as related to its property $Q_{X_i}$) depends on $x_i$ and $f_i$, similarly as $Q_X$ depends on $x$ and $f$ (Eq. 15). Here $x_i$ is a neurobiological quantity localized in the brain, albeit diffusely. However, the fitness component $f_i$ is not a clearly localized quantity. It could get contributions from anything in the world, as long as those things belong to $w$ (see Eqs. 1 and 4). This means that the cause $X_i$ is not well localized either. Moreover, only $x_i$, not $f_i$, is directly controlled by the individual. Hence, the individual does not fully own the cause $X_i$. In conclusion, the cause $X_i$ does not seem to be a good candidate for consciousness: although it is distinct, it is neither clearly localized to, nor fully owned by, the individual. It cannot be ruled out that the individual might weakly sense this cause (weakly because it is not localized), but it is not directly communicable. Thus, it is different from how consciousness is conventionally understood (as, in principle, reportable).

The situation is quite different for the cause $\bar{X}_i$. That cause (as related to its property $Q_{\bar{X}_i}$) depends on $x_i$ and $X_i\hat{u}_i$ (Eq. 17). Here, all quantities involved ($x_i$, $X_i$, and $\hat{u}_i$) are localized in the neurobiology of the brain. Such localization is somewhat diffuse, as these quantities are distributed over different parts of the brain (such as in Fig. 9b), but they do not extend beyond the brain. Moreover, all quantities are part of the individual. Therefore, the cause $\bar{X}_i$ is, in some sense, owned by the individual. In conclusion, the cause $\bar{X}_i$ is a good candidate for consciousness, because of spatially confined localization and clear ownership. Below I will argue that it is plausible that $\bar{X}_i$ is *felt* by the individual.

According to Figs. 6-8, the operator $\bar{X}_i$ can be realized by a switched use of $X_i$ in a feedback loop. Each time when $\bar{X}_i$ is realized in this way, it is a distinct cause (of the $f_+$ that arises from producing $\hat{u}_i$, on average and possibly only in the future). Thus, $\bar{X}_i$ exists distinctly at that time. This distinct existence is transient, because $\bar{X}_i$ is produced in a pulsed manner. Moreover, the nature of $\bar{X}_i$ changes, because it is a nonstationary function of time. Finally, the distinct existence of $\bar{X}_i$ is produced by the fact that it depends on modulated noise, rather than being solely produced by material components. Nevertheless, the cause $\bar{X}_i$ exists in a real, literal sense. What we have here, then, is a real, spatially localized cause that briefly appears in one's head, at the rate of switching to stage 2 in the model of Fig. 8. If this real, spatially localized cause would correspond to a material object, then there would be little doubt that its pulsed presence in the brain would be sensed by the individual. Although the cause here is not purely material, but transient and produced by a rather special form of causation, it is plausible that it is sensed as well: it is real, distinct, spatially localized, and it has material effects. Its pulsed presence in one's head is conjectured here to produce a sensation that can be equated to feeling conscious in general, as well as to feeling aware of the specific content associated with $\bar{X}_i$.

## 10 Discussion and prospect

The theory explained above proposes that consciousness is equivalent to sensing the distinct, partly autonomous cause that arises when an individual prepares to communicate (possibly for internal use) estimated aspects of its own evolutionary fitness. Moreover, it proposes that such preparation is carried out by the second stage of a dual use of the thalamocortical feedback loop. Whereas the first stage merely produces estimates, the second stage inverts them. Only if the latter occurs, consciousness results. The theory will be discussed below, first, with respect to related approaches in the neurosciences, and, second, with respect to how it can be tested empirically.

The theory shares characteristics with existing theories of consciousness, particularly ones closely aligned with neurobiological findings. Having two stages, a nonconscious and conscious one, is similar to what is proposed in the global neuronal workspace theory (Dehaene et al. 2003; Dehaene 2014; Dehaene et al. 2017). If one would define a workspace (e.g., in Fig. 9), it would be one jointly produced by thalamus and cortex. Whereas Dehaene and colleagues localize the workspace primarily in the cortex, related work on blackboard theory (Mumford 1991) localizes it primarily in the thalamus. Having two stages is also a characteristic of the recurrent processing theory (Lamme and Roelfsema 2000; Lamme 2006; van Gaal and Lamme 2012). Recurrent processing is presented there as a cortical phenomenon, without an explicit role for the thalamus. But the latter may well be consistent with the observations on which that theory is based.

The current proposal locates the content of consciousness in the components $X_i$, which are mainly produced in the cortex. This content also exists in stage 1, but without being conscious at that time. Only in stage 2, when the $X_i$s are inverted, such content becomes conscious. But it would be a mistake to locate consciousness purely in cortex (see also Merker 2007). The process of inversion requires the subtractive feedback that is conjectured here to be located in TRN and dorsal thalamus. Thus, without the thalamus there could be no consciousness. Moreover, if more and more of the cortex would become dysfunctional, more and more content would be lost, but not consciousness itself (as long as at least some feedback loops remain).

The central role of the thalamus and the thalamocortical loop for consciousness has been noted before, such as in the dynamic core theory of Edelman and Tononi (2000). An elaboration of this theory specifically proposes that integration and differentiation of information produce consciousness (e.g., Oizumi et al. 2014; Tononi et al. 2016b). Integration and differentiation of content are also features of the theory proposed here, being produced by $X$. Anything pertaining to $X$ is automatically integrated, because its effect depends crucially on $x$, a scalar. At the same time, differentiation is produced by which sets of components $X_i$ are active at any particular point in time. Thus, integration and differentiation are ascribed here not uniquely to consciousness, but already to the nonconscious production of $X$. Consciousness is ascribed here to the production of sets of $\bar{X}_i$s, which all depend, ultimately, on $x$ as well. Hence, the current theory is consistent with the fact that consciousness is experienced as both integrated and differentiated.

Ward (2011, 2013) provides a wealth of arguments for why the thalamus is essential for consciousness. One interesting observation is that different anesthetics modify activities in various parts of the brain, but with a common focus on the thalamus (Alkire and Miller 2005; but see also Alkire et al. 2008; Tononi et al. 2016a). Breaking or disturbing the feedback loop that inverts $X_i$ (in the lower part of Fig. 9b) is indeed expected to prevent consciousness. Another interesting point discussed by Ward is that damage to the thalamus or connecting axons is strongly associated with the unresponsive wakefulness syndrome (vegetative state, Jennett et al. 2001; for reviews see Schiff 2010; Giacino et al. 2014). Again, this is consistent with the proposal of Fig. 9b. It should be noted that just removing or disturbing stage 2 would already prevent consciousness, even if stage 1 is left relatively undisturbed. Such subtle changes in circuit performance at fast timescales may not be easily resolvable by some of the available experimental techniques (which may explain conflicting results on the role of the thalamus in consciousness, which are discussed by Koch et al. 2016).

A specific role for the TRN in generating consciousness is proposed by Min (2010). That proposal is in the tradition of relating consciousness to the presence of the synchronized oscillations that one can find in thalamus and cortex (Crick and Koch 1990; Llinas et al. 1998; Singer 2001; Engel and Fries 2016). The current theory does not directly address such oscillations. Nevertheless, there are two distributed processes that could benefit from the synchronization that oscillations can provide. The first process is the production of $X$, across cortex and subcortical nuclei. The second is the proposed two-stage functioning of the thalamocortical loop (Fig. 9b). Switching between the two stages requires considerable synchronization, first, between thalamus and cortex, and, second, across the cortex for producing $X$. Thus, synchronization is then a prerequisite for producing consciousness.

The interaction of TRN with dorsal thalamus has been conjectured to optimize perception (Harth et al. 1987) and to affect attention (the searchlight hypothesis of Crick 1984). Indeed, there is now direct evidence that it affects attention in mice (Wimmer et al. 2015). Attention and consciousness are related, but they are not the same (Koch and Tsuchiya 2007; van Boxtel et al. 2010). In particular, attention does not require consciousness. The current theory does not conflict with an attentional role for the TRN and thalamus, because they are sufficiently complex to carry out various tasks simultaneously. Attention may in fact have evolved before consciousness, such as through inhibitory connections from TRN to dorsal thalamus. This could then have facilitated the subsequent evolution of the feedback circuit that is proposed here for inverting $X_i$, through modest additions to and modifications of existing circuits.

Interestingly, the specific thalamocortical structure of Fig. 9a appears to be best developed in mammals (Butler 2008), but similar structures are present in birds and reptiles as well, to a lesser extent. It is absent in fishes and amphibians. This is consistent with the fact that the presence of consciousness seems most easily ascribable to mammals, birds, and some reptilian species (according to the theory, the capability to make distress calls suggests consciousness; see also Panksepp and Panksepp 2013). However, the number of species that have been investigated anatomically is limited, which means that there may be more variation among species than is presently known. Moreover, Butler (2008) notes that in species lacking a TRN some of its functions might be realized by other pathways (such as the loop through the striatum, which is always present).

The theory depends on producing internal estimates $x_i$ of partly external quantities $f_i$. This is similar to theories of the brain that depend on internal models of external reality (if we ignore, for the moment, that consciousness is produced here not by estimating, but by inverting estimates). Such theories propose that the prime evolved function of the brain is to predict external and internal states, which can be formulated in terms of predictive coding, maximization of mutual information, and minimizing variational free energy (e.g., Friston 2010; Clark 2013; Seth 2015). Relatedly, the good regulator theorem (Conant and Ashby 1970) states that the proper way to regulate a system is to utilize a good model of it. It is quite possible that much of the brain's functioning can be understood in this way. However, this applies to response strategies that have been evolved or learned, or that are readily learnable. In Section 3 such strategies were acknowledged as an established baseline on top of which $X$ works. $X$ is not about regulating, but about producing novel, unanticipated structures within the brain. It concerns non-ergodic change, not homeostasis. Importantly, $X$ produces unity in the organism. This is so because $X$ is a distinct cause that depends crucially on $x$, which is a scalar that estimates another scalar, $f$. The fundamental reason why the latter is a scalar, and is thus unitary, is that organisms survive and reproduce as wholes, not as parts. In contrast, specialized (i.e., ad hoc) internal models of external reality cannot provide unity, because external reality is not unitary by itself (i.e., without being treated by an organism as an integrated part of its $f$, through $x$ and the stochastic mechanism by which $x$ produces effects). Specialized internal models may still evolve whenever they increase fitness $f$ in a direct way, but neither such models nor their inverses would produce distinct causes. Hence, they cannot produce a cause of the kind that is conjectured here to be felt.

The theory presented here is clearly conjectural. The question is then how it can be tested empirically. Although the model presented in Figs. 7 and 8 is abstract, it is formulated in such a way that it can be translated into a detailed neurophysiological model, at least in principle. A first, still rather coarse attempt is presented in Section 8 and Fig. 9. The theory is of a kind that can, ultimately, only be rigorously tested by a combination of detailed modelling and measurements. It predicts that neurophysiological models exist that simultaneously fulfil three conditions: first, they explain and predict the corresponding neurophysiological measurements, second, they conform to the overall dynamical structure of the model in Figs. 7 and 8, and third, the properties of the model's switched dynamics should correspond to those of subjective experience. Such detailed and brain-wide neurophysiological models may only be possible in the long run. Therefore, I will make a few suggestions here for more preliminary empirical assessments of the theory's viability.

A major prediction that should produce empirical observables is the continual switching of the thalamocortical loop between two states (roughly at 10 Hz in primate brains). One state produces only components $X_i$ (corresponding, roughly, to unaware representations) and the other produces components $\bar{X}_i$ (corresponding to aware representations). By empirically interfering with the second cycle, it should be possible to suppress consciousness without interfering much with the establishment of implicit representations. An experiment along these lines would require monitoring the phase of the state-switching, and precisely timed interference.

A further major prediction is that when a sender and receiver are engaged in a consciously perceived (nonverbal) dialogue (as in Fig. 7), their thalamocortical loops must become synchronized. Randomly interfering with the timing of the state-switching in either of the two communicating partners should prevent synchronization and suppress consciously perceived dialogue. The latter should also happen when the disturbed dialogue is merely imagined or rehearsed (one possible mode of operation of Fig. 8). Moreover, the awareness in a receiver may be manipulated, in a predictable way, by the precise timing and content of the communications by a simulated sender that is in dialogue with the receiver.

The internal dialogue of Fig. 8 suggests another way to interfere with consciousness. Buffers are required for the two stages (forward $X_i$ and inverted $X_i$) to work together. If either of the buffers $x_i(t)$ and $\hat{u}_i(t)$ can be localized, interfering with their loading or their retention time should produce malfunctions of the loop in predictable ways.

If the interpretation of Fig. 9 is correct, then the TRN should have a major role in regulating consciousness. The role of the TRN for attention in mice has recently been studied in a range of sophisticated experiments (e.g., Wimmer et al. 2015). Although attention and consciousness are not the same, they are related (as is briefly mentioned above). It may be possible to adjust the experimental protocols of these experiments in such a way that they provide information on the function of the TRN for consciousness rather than for attention.

Finally, the theory depends on the conjecture that there is an $X$-process that produces an internal estimate of an organism's inclusive fitness, and that the value of this estimate then modulates the variability of behavioural dispositions. Although $X$ does not produce consciousness by itself, the fact that it is a distinct cause is used in the theory to argue that each $\bar{X}_i$ is a distinct cause as well, resulting in the feeling of consciousness. Testing for the presence of $X$ and its effects requires, ultimately, detailed and brain-wide modelling as well. A tentative test may involve manipulating fitness (or at least apparent fitness) and observing the resulting variability of behaviour or behavioural dispositions.